\documentclass{aa}  

\usepackage{amsmath}
\usepackage{graphicx}
\usepackage{hyperref}
\usepackage{multirow}
\usepackage{comment}
\usepackage{caption}
\usepackage{subcaption}
\usepackage{graphicx}
\usepackage{txfonts}
\usepackage{float}
\begin{document}

   \title{Pre-computed aerosol extinction, scattering and asymmetry grids for scalable atmospheric retrievals}

   \author{M. Voyer\inst{1}
          \and
          Q. Changeat\inst{2}
          }

   \institute{Université Paris Cité, Université Paris-Saclay, CEA, CNRS, AIM, F-91191 Gif-sur-Yvette, France\\
              \email{mael.voyer@cea.fr}
         \and
             Kapteyn Institute, University of Groningen, 9747 AD Groningen, NL.\\
             }

   \date{Received xx, xx; accepted xx xx, xx}

 
  \abstract{The unprecedented wavelength coverage and sensitivity of the James Webb Space Telescope (JWST) permits to measure the absorption features of a wide range of condensate species from Silicates to Titan tholins. Atmospheric retrievals are uniquely suited to analyse these datasets and characterize the aerosols present in exoplanet atmospheres. However, including the optical properties of condensed particles within retrieval frameworks remains computationally expensive, limiting our ability to fully exploit JWST observations.}
   {In this work, we improve the computational efficiency and scaling behavior of aerosol models in atmospheric retrievals, enabling in-depth studies including multiple condensate species within practical time scales.}
   {Rather than computing the aerosol Mie coefficients for each sampled model, we pre-compute extinction efficiency (${\mathrm{Q_{ext}}}$), scattering efficiency ($\mathrm{Q_{scat}}$) and asymmetry parameter ($\mathrm{g}$) grids for seven condensate species relevant in exoplanet atmospheres ($\mathrm{Mg_2SiO_4\, amorph\,sol-gel}$, $\mathrm{MgSiO_{3}\, amorph\,glass}$, $\mathrm{MgSiO_{3}\,amorph\,sol-gel}$, $\mathrm{SiO_{2}\,alpha}$, $\mathrm{SiO_{2}\,amorph}$, $\mathrm{SiO}$ and Titan tholins). During retrievals, the relevant values are obtained via linear interpolation between grid points which drastically reduces computation times.}
   {The pre-computed ${\mathrm{Q_{ext}}}$ grids significantly reduce computation time between 1.4 and 17 times with negligible differences on the retrieved parameters. They also scale effortlessly with the number of aerosol species while maintaining the accuracy of cloud models. Thereby enabling more complex retrievals as well as broader population studies without increasing the overall error budget. The ${\mathrm{Q_{ext}}}$, $\mathrm{Q_{scat}}$ and $\mathrm{g}$ grids are freely available on Zenodo as well as a public \textsc{TauREx} plugin---\textsc{TauREx-PCQ}---that utilize them.}
   {Whether in emission, transit or phase curve, clouds play a major role in the spectra of exoplanets, brown dwarfs or proto-planetary disks, with the piling JWST observation and the future ARIEL space telescope it is essential to improve retrieval frameworks to handle both high information content and population level datasets. With their speed ups, the aerosol grids presented in this paper are a significant step forward in that direction.}


   \maketitle


\section{Introduction}

The launch of the James Webb Space Telescope (JWST) has unlocked characterization of exo-atmospheres, improving by orders of magnitude the spectral resolution and wavelength coverage of exoplanet spectra \citep{rigby_SciencePerformanceJWSTCharacterized_2023}. This drastic increase in data information content has opened the doors to new scientific breakthroughs including isotopologue detections, previously unseen molecules, characterization of complex thermal structures, identification of clouds, and spatial mapping \citep{barrado_15NH3AtmosphereCoolBrown_2023,kuhnle_WaterDepletion15NH3Atmosphere_2025,tsai_PhotochemicallyProducedSO2Atmosphere_2023,matthews_HCNC2H2AtmosphereT85+T9_2025, voyer_MIRILRSSpectrumColdExoplanet_2025,molliere_EvidenceSiOCloudNucleation_2025,changeat_CloudHazeParameterizationAtmospheric_2025,murphy_EvidenceMorningtoeveningLimbAsymmetry_2024}. With increasing data complexity and new physico-chemical processes to constrain, models now have to incorporate many more degrees of freedom. Atmospheric retrievals---the most utilized statistical inversion technique for exo-atmospheric data---typically require $>20$ free parameters \citep{gandhi_JWSTMeasurements13C18O_2023,kothari_ProbingHeightsDepthsDwarf_2024,matthews_HCNC2H2AtmosphereT85+T9_2025,changeat_CloudHazeParameterizationAtmospheric_2025}. A significant fraction of this increase in complexity (i.e. an increase in computational cost) relates to the modeling of aerosols species, which are now accessible and need to be modeled using non grey prescriptions, for instance see e.g., VHS-1256\,b or WASP-107\,b \citep{miles_JWSTEarlyreleaseScienceProgram_2023,dyrek_SO2SilicateCloudsNo_2024}. To model clouds and hazes, modern retrieval frameworks like \textsc{petitRADTRANS} or \textsc{TauREx} use optical constants measured in laboratory experiments. These input data are then used to model the absorption features of e.g., iron, silicates, titan hazes or water \citep{molliere_PetitRADTRANSPythonRadiativeTransfer_2019,changeat_CloudHazeParameterizationAtmospheric_2025}. The method, however, requires the computation of the aerosol wavelength-dependent extinction using Mie theory \citep{bohren_absorption_1983}. Solving these equations for each retrieval sample requires significant computational time, sometimes even being the main computational bottleneck \citep{sumlin_RetrievingAerosolComplexRefractive_2018,changeat_CloudHazeParameterizationAtmospheric_2025}. 
Parallel to the developments of these data driven methods (i.e., free  retrievals), self-consistent models have also improved drastically now including complex clouds formation, nucleation growth, gas phase modeling and $3$D global climate models \citep{min_ARCiSFrameworkExoplanetAtmospheres_2020,ma_YunMaEnablingSpectralRetrievals_2023,helling_ExoplanetWeatherClimateRegimes_2023}. first principle approaches also employ the wavelength dependent scattering coefficients and asymmetry parameters to fully model the aerosols behaviors. The opacity of cloud particles have been heavily scrutinized where deviations from filled spheres (e.g., fractal or porous particles) has been shown to induce significant differences in extinction \citep{ohno_CloudsFluffyAggregatesHow_2020,kiefer_WhyHeterogeneousCloudParticles_2024}. However, the computational cost to include these complex models has so far mostly prevented their inclusion in atmospheric retrievals \citep{ma_YunMaEnablingSpectralRetrievals_2023}. To unlock multi-instrument, population level, retrieval studies with JWST and with the future ESA-Ariel telescope\footnote{Ariel will observe thousands of exoplanets, producing up to three transmission spectra a day, with wavelengths from $0.7$ to $7.8$ microns that are sensitive to haze and clouds \citep{tinetti_ariel_2021}.}, the computation of aerosol extinction must be optimized.
Here, we focus our study on homogeneous spherical  particles to build fast and scalable haze and clouds radiative models to analyze populations of exo-atmospheres with  atmospheric retrievals. In this paper, we aim to improve retrieval speed by removing on-the-fly Mie calculations in clouds models, instead relying on pre-computed extinction, scattering and asymmetry coefficients. This method is somewhat similar to what is employed for molecular cross-sections \citep{chubb_ExoMolOPDatabaseCrossSections_2021} and to the procedure presented in \cite{batalha_condensation_2025} for the \textsc{virga-v1} clouds models. However, the technique is not commonly employed by the retrieval community, and no sensitivity study ensuring numerical convergence has been performed so far. Here, we construct a new \textsc{TauREx} plugin: \textsc{TauREx-PCQ}, offering this new feature to the widest community in an open-source flexible framework, and we perform rigorous benchmark to validate all the assumptions baked in the pre-computation of the extinction grids. Section \ref{sec:2} describes how the different grids are built and validated, while Section \ref{sec:3} demonstrates performance improvements for four different test cases. In that last section, we explicitly benchmark pre-computed extinction coefficient retrievals (hereafter PCQ retrievals) against retrievals using on-the-fly Mie computations.

\section{Methodology}\label{sec:2}

Recently, retrieval codes have included Mie theory to model the absorption of clouds in exo-atmospheres \cite[see e.g., ][]{molliere_PetitRADTRANSPythonRadiativeTransfer_2019, changeat_CloudHazeParameterizationAtmospheric_2025}. Mie theory is non-linear with respect to particle size (noted $a$), but using fine grid sampling can enable linearization for atmospheric retrieval. Here, we implement such an approximation and estimate its impact on the induced errors of atmospheric retrievals (i.e. we compare the recovered posterior distributions).
As our primary goal is to improve the speed and scalability of free retrievals we first focus on $\rm{Q_{ext}}$. Then, in section \ref{sec:Qscat_g} we generalize our methodology to $\mathrm{Q_{scat}}$ and $\mathrm{g}$ which are used in broader applications. To compute haze and cloud absorption, one needs to calculate the extinction of a single particle as a function of its size and wavelength (noted $\lambda$). We use the open-source python library \textsc{PyMieScatt} \citep{sumlin_RetrievingAerosolComplexRefractive_2018}, packaged in the \textsc{TauREx-PyMieScatt} \citep{changeat_CloudHazeParameterizationAtmospheric_2025} plugin to compute the absorption of aerosol particles. \textsc{PyMieScatt} performs on-the-fly inversion of Lorenz-Mie theory equations to compute the aerosol's extinction coefficient ($\rm{Q_{ext}}$) of a particle at size $a$ and wavelength $\lambda$:

\begin{figure}[ht!]
    \centering
    \includegraphics[width=\columnwidth]{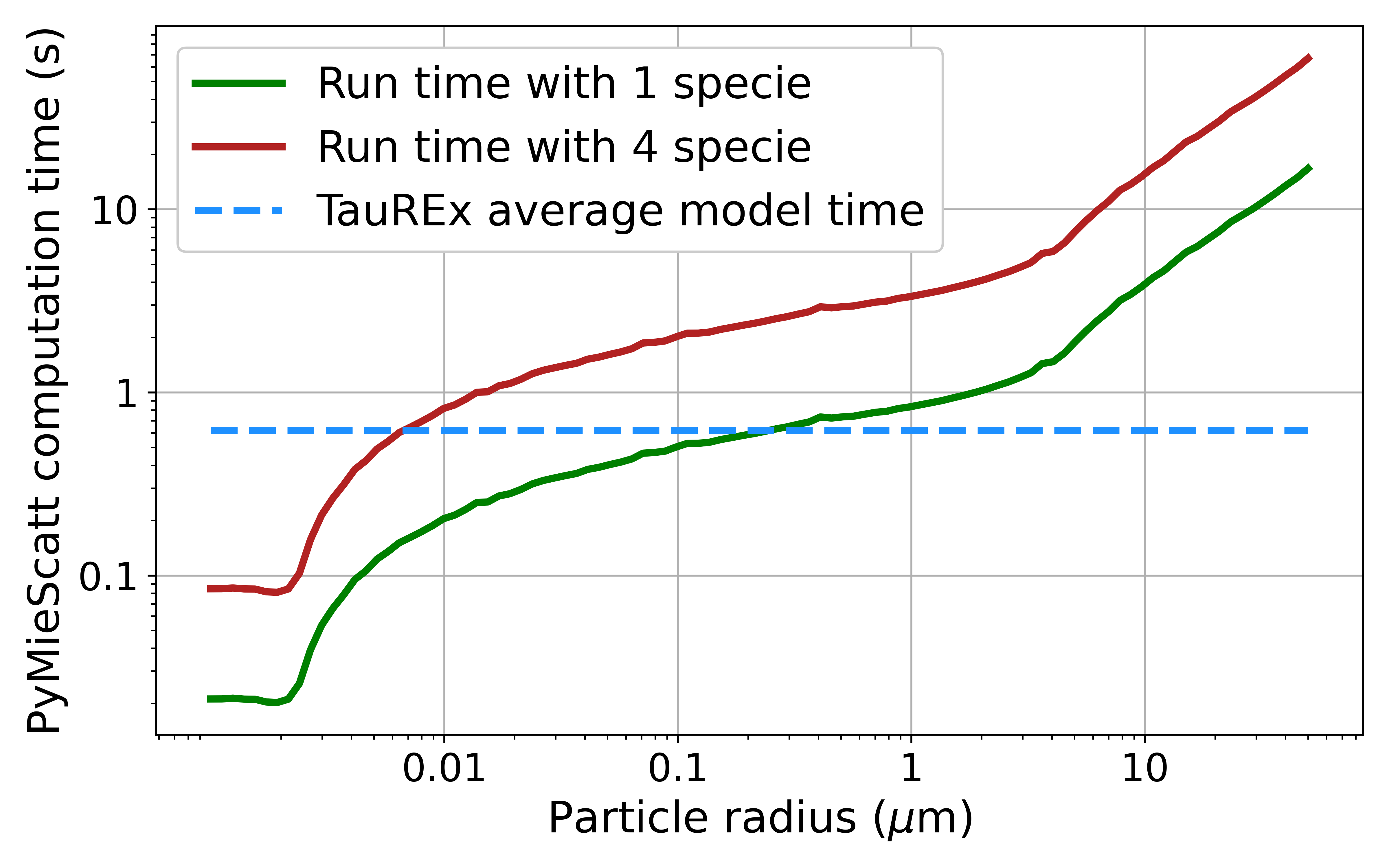}
    \caption{PyMieScatt computation time for the extinction coefficient of a Titan tholin spherical particle versus the particle's size. For each particle size, the $\rm{Q_{ext}}$ are computed on the full wavelength grid of the optical constants (see section \ref{sec:optical_cste}).The green and red lines represent the run time of a model with \textsc{PyMieScatt} with one and four clouds species respectively. The time displayed are the median times for 10 iterations to avoid outliers. The dashed blue line is the average model time for a cloud-free \textsc{TauREx} forward model with $100$ layers \citep{al-refaie_TauREx3FastDynamic_2021}.}
    \label{Fig:MieTime}
\end{figure}

\begin{equation}
    \mathrm{Q}_{\text{ext}}(a, \lambda) = \frac{2}{x^2} \sum_{n=1}^{n_{max}} (2n + 1) \operatorname{Re} \left( a_n(m(\lambda), x) + b_n(m(\lambda), x) \right)
\label{eq:1}
\end{equation}

with $x = \frac{2\pi \rm{a}}{\lambda}$ and the Mie coefficients $a_n$ and $b_n$ being defined with the complex optical constants $m(\lambda) = n(\lambda) - i k(\lambda)$. The computational time needed for these calculation increases rapidly with radius of the particle (see Fig. \ref{Fig:MieTime}). This is because the  number of multipole order required for a correct approximation $n_{max}$ is given by :
$$\rm{n_{max} = x + 4.3\cdot x^{1/3}}.$$
For a more complete description, please refer to \cite{bohren_absorption_1983} and \cite{kitzmann_optical_2018}. Fig. \ref{Fig:MieTime} also shows the average time for a cloud-free forward model with TauREx \citep{al-refaie_TauREx3FastDynamic_2021}: the forward model computing time becomes dominated by cloud calculations when $a > 0.3\,\mu$m when an single cloud species is considered. However, in the JWST era, atmospheric retrievals often require more than one cloud species to explain the observed data \citep{molliere_EvidenceSiOCloudNucleation_2025,changeat_CloudHazeParameterizationAtmospheric_2025}. When including four aerosols the forward model computation is slowed down by the Mie calculation as soon as $a >0.008\,\mu$m (see Fig. \ref{Fig:MieTime}). Additionally, with Nested Sampling retrievals---the most commonly used by the exoplanet community---the parameter space exploration is synchronized between MPI processes every $N$ live point evaluations. Optimal scaling performances are therefore obtained when all forward model evaluations take the same amount of time, otherwise faster processes halt and wait for synchronization. With Mie calculations poor scaling with $a$, the progression of the algorithm always scales with the largest evaluated $a$, wasting significant computing resources. To avoid slowing down retrievals with these computations, a better scaling model is required.  To do so, we pre-compute $\rm{Q_{ext}}$ for different molecules and for a range of $a$. With this method, estimating a $\rm{Q_{ext}}$ can be done by interpolating between the closest known values. However, since radiative transfer is highly non-linear, even seemingly negligible errors on extinction coefficient can produce strong biases. We therefore conduct rigorous tests to build grids at {\it safe}, numerically adapted resolutions. We utilize the retrieval framework \textsc{TauREx3}\footnote{\url{https://taurex.space/}} and its recently published \textsc{TauREx-PyMieScatt}\footnote{\textsc{TauREx-PyMieScatt} is available at :\url{https://github.com/groningen-exoatmospheres/taurex-pymiescatt}} plugin in \cite{changeat_CloudHazeParameterizationAtmospheric_2025} to compare between retrievals using on-the-fly Mie computation or linearized extinction coefficient \citep{al-refaie_TauREx3FastDynamic_2021}.

\subsection{Optical constants}\label{sec:optical_cste}
\begin{figure}[ht!]
    \centering
    \includegraphics[width=\columnwidth]{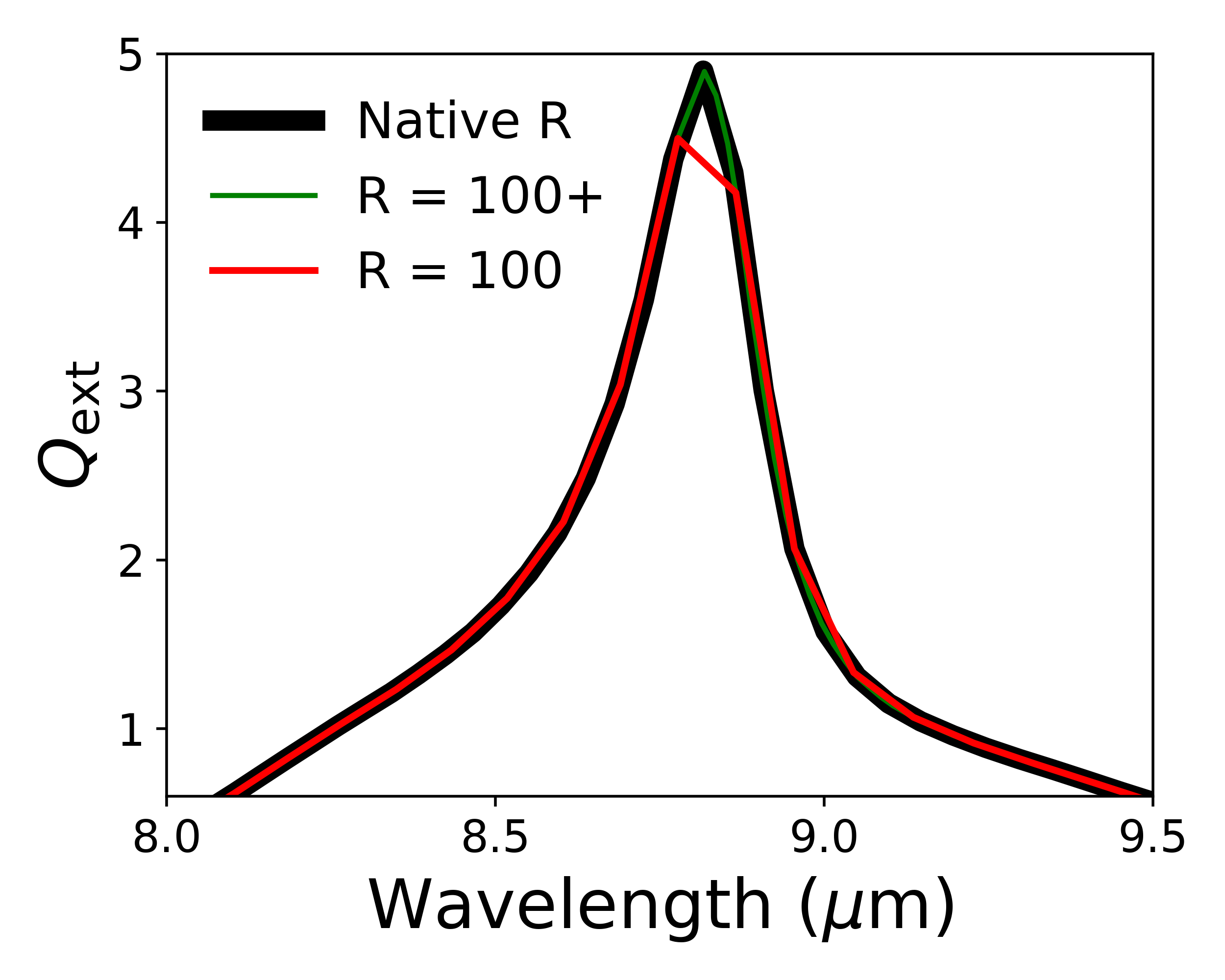}
    \caption{Extinction coefficient for a $700$ nm SiO$_2$ amorph particle versus wavelength. The black, red and green lines respectively shows the $\mathrm{Q}_{\text{ext}}$ computed using optical constants at native resolution, at a resolution of $100$ and at a global resolution of 100 but with a local 500 resolution for key features.}
    \label{fig:cindex}
\end{figure}

To compute the extinction, scattering and asymmetry coefficients, one requires the complex optical constants of the particles $m(\lambda)$ (see Eq. \ref{eq:1}). These constants are measured experimentally in laboratories using Kramers-Kroning analysis and Lorenz oscillator fit method on the reflection spectra of the powdered compound \citep{jager_StepsInterstellarSilicateMineralogy_2003}. The variations and features of $m(\lambda)$ create the absorption features of the aerosols (see Fig. \ref{fig:cindex}). Therefore, to accurately model cloud and haze absorption bands, the optical indexes must be measured with sufficient spectral resolution \citep{bohren_absorption_1983, kitzmann_optical_2018}. In the literature, however, the spectral resolution and wavelength coverage of available optical indexes varies \citep{
jager_StepsInterstellarSilicateMineralogy_2003,dorschner_StepsInterstellarSilicateMineralogy_1995,zeidler_OpticalConstantsRefractoryOxides_2013,henning_LowtemperatureInfraredPropertiesCosmic_1997,palik_handbook_1985,khare_OpticalConstantsOrganicTholins_1984}. Other studies like \cite{batalha_condensation_2025} have adopted a common wavelength grid, interpolating and/or extrapolating optical indexes between $0.3$ to $300$ microns. They use a grid with a varying resolution from $2.5$ to $250$ but recommend a resolution of $300$ to interpret JWST datasets following \cite{grant_JWSTTSTDREAMSQuartzClouds_2023}.

\begin{figure*}[ht!]
   \centering
    \includegraphics[width=0.99\textwidth]{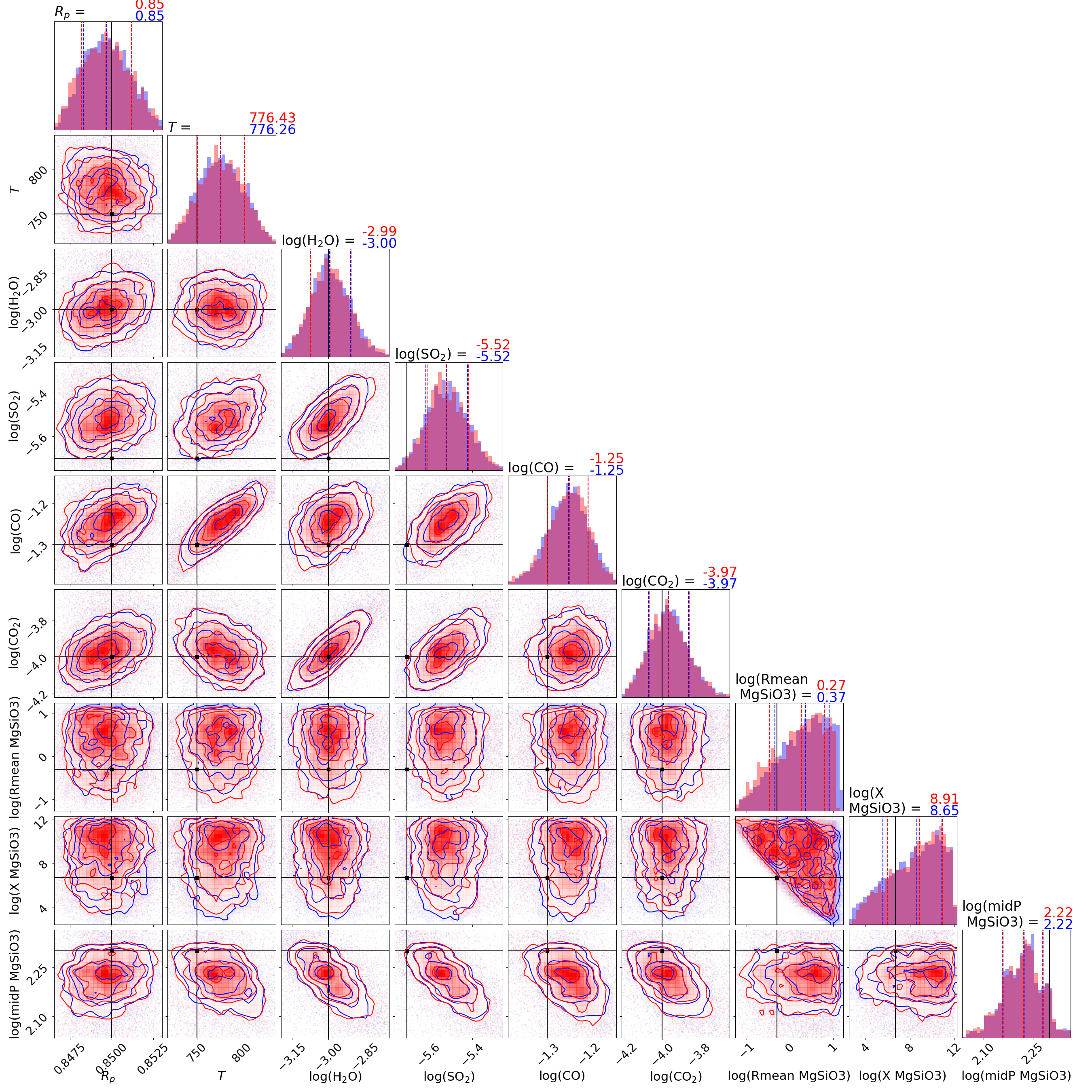}
   \caption{Posterior distributions for WASP 107\,b inspired self-retrieval, the \textsc{PyMieScatt} and \textsc{TauREx-PCQ} are shown in blue and red respectively. The median value of each retrieved parameter is shown on top of the corresponding histogram. The truth values are displayed in black.}
              \label{Fig:wasp}%
\end{figure*}

In \cite{changeat_CloudHazeParameterizationAtmospheric_2025}, the authors interpolate the optical indexes at a resolution $100$ for all the aerosols in \textsc{TauREx-PyMieScatt}. In many cases, this resolution is well suited for Mie computations and pre-computed $\rm{Q_{ext}}$ grids because of the relatively low number of points which limits computation time and memory cost. On the interval $\lambda \in [0.3 - 50]\, \mu$m---i.e. the wavelength range of \textsc{EXOMOL}'s cross-sections---only $517$ wavelength points are needed. In some cases, such as $\rm{MgSiO_3}$, the optical constants need to be interpolated because they are experimentally measured at a lower spectral resolution than $100$. In this case, the interpolation, and hence our knowledge of the optical constants, becomes the dominant error factor. For some species presenting sharp features, such as SiO$_2$ amorph (or SiO, SiO$_2\,\alpha$), a higher resolution is needed (and measured): the $100$ resolution does not suffice to accurately interpolate the variations of $n(\lambda)$ and $k(\lambda)$ as shown in Fig. \ref{fig:cindex}. For a particle of SiO$_2$ amorph with $a = 700$\,nm, an interpolation of $n(\lambda)$ and $k(\lambda)$ at resolution $100$ induces a relative error of $\sim 4\%$ in the tip of the Si -- O stretch band. While such level of uncertainty does not affect the retrieval outcomes of transit spectroscopy retrievals (see Fig. \ref{Fig:wasp}), they are sufficient to significantly bias the retrieval of directly imaged objects (see Fig. \ref{Fig:cornerResolution}). To correct for this bias, we combine the wavelength grid at resolution $100$ from \cite{changeat_CloudHazeParameterizationAtmospheric_2025} with local, higher resolution grids (between $500$ and $1000$) in manually selected, key wavelength region to produce smooth $\mathrm{Q}_{\text{ext}}$ features. This method only adds between $30$ to $100$ points to the wavelength grid depending on the species and much better suited for high information content retrieval analysis (see Fig. \ref{Fig:cornerResolution}).

\subsection{Radius grid and $\rm{Q_{ext}}$ computations}\label{sec:radius_grid}

We follow Mie theory from \citep{bohren_absorption_1983}. In exo-atmospheric studies, we seek to constrain the cloud particle mass mixing ratio, vertical extend and particle radius $a$. Often, codes like \textsc{TauREx} or \textsc{PetitRADTRANS}, retrieve $a$ as a free parameter from the cloud absorption properties in the spectrum (i.e., scattering slope and aerosol features). Since cloud layers display a distribution of particles rather than single-sized particles, a few evaluation of $\rm{Q_{ext}}$ at different $a$ are needed for each model. Pre-computing $\rm{Q_{ext}}$ can save significant computing time. Here, we pre-compute $\rm{Q_{ext}}$ for all the relevant species in Table \ref{tab:placeholder}. We describe the procedure for Titan's tholin but the same steps are applied to all molecules. Based on recent studies, aerosol particle sizes in planetary and sub-stellar atmospheres typically range from about $10$ nm to $10$ µm \citep{gao_AerosolCompositionHotGiant_2020,teinturier_CloudsDriverVariabilityColour_2025}. To capture a conservative range of plausible grains, we adopt $a \in [0.001, 30]\,\mu$m and compute $\rm{Q_{ext}}$ with \textsc{PyMieScatt} \citep{sumlin_RetrievingAerosolComplexRefractive_2018} for the wavelengths $\lambda\in[0.3, 50]\,\mathrm{\mu m}$. We chose these wavelengths limits to match those of the EXOMOL's cross sections as they are intended to be used together in retrievals \citep{tennyson_exomol_2016}. 

\begin{table}[ht!]
\caption{Aerosol species now available on the Zenondo repository with the respective sizes of the associated grids and reference of the optical indexes.}
\tiny
    \centering
    \begin{tabular}{c|c|c}
 Aerosol name&Grid size &Reference\\\hline\hline
         $\rm{Mg_2SiO_4}\, amorp\,sol-gel$& $264$Mb & \cite{jager_StepsInterstellarSilicateMineralogy_2003} \\\hline
         $\rm{MgSiO_{3}\, amorph\,glass}$& $317$ Mb & \cite{dorschner_StepsInterstellarSilicateMineralogy_1995} \\\hline
         $\rm{MgSiO_{3}\,amorph\,sol-gel}$& $265$Mb & \cite{jager_StepsInterstellarSilicateMineralogy_2003} \\\hline
         $\rm{SiO_{2}\,alpha}$& $387$Mb & \cite{zeidler_OpticalConstantsRefractoryOxides_2013} \\\hline
 $\rm{SiO_{2}\,amorph}$&$258$Mb & \cite{henning_LowtemperatureInfraredPropertiesCosmic_1997} \\\hline
 $\rm{SiO}$&$179$Mb & Philipp in \cite{palik_handbook_1985} \\\hline
 $\rm{Titan\,tholins}$&$240$Mb & \cite{khare_OpticalConstantsOrganicTholins_1984}  \\ \hline
    \end{tabular}
    \tablefoot{ The optical constants of all listed species are measured between $0.3$ and $50$ microns at the least. "amorph" stands for the amorphous form of these condensates while "sol-gel" refers to the chemical process used to form the particles (see \cite{jager_StepsInterstellarSilicateMineralogy_2003} for details). The term "alpha" designs a specific crystalline structure of $\rm{SiO_2}$.}
    \label{tab:placeholder}
\end{table}

We do not use the same grid spacing in $a$ for all the species. Instead, we adopt the smallest (in hard-drive space) non-uniform grid that ensures a target accuracy for $\rm{Q_{ext}}$ (see Table \ref{tab:placeholder}), which is different for each species. In this work, the accuracy is defined by the relative error to the exact computation with PyMieScatt, and it is set to $\epsilon = 10^{-4}$. The following paragraph describes how the radius grids are built. 

\begin{table*}[h!]
\caption{Overview of the test cases used to validate our approach.}
    \centering
    \begin{tabular}{c|c|c|c|c}
        Planet & Method & Facility & Instrument & Cloud species\\\hline\hline
        $\mathrm{WASP-107\,b}$ & Transit & JWST & NIRISS, NIRSpec $\mathrm{G395H}$ and MIRI LRS & MgSiO$_3$ amorph glass \\\hline
        $\mathrm{HD\,189733\,b}$ & Transit & ARIEL & FGS $\&$ AIRS & MgSiO$_3$ amorph glass \\\hline
        $\mathrm{2MASS\,J2236+4751\,b}$ & Direct spectroscopy & JWST & NIRSpec $\mathrm{G395M}$ \& MIRI LRS & MgSiO$_3$ amorph glass  \\\hline
        $\mathrm{GJ\,436\,b}$ & Transit & ARIEL & FGS $\&$ AIRS & Titan Tholin\\
    \end{tabular}
    \label{tab:cases}
\end{table*}

\begin{figure}[ht!]
    \centering
    \includegraphics[width=\columnwidth]{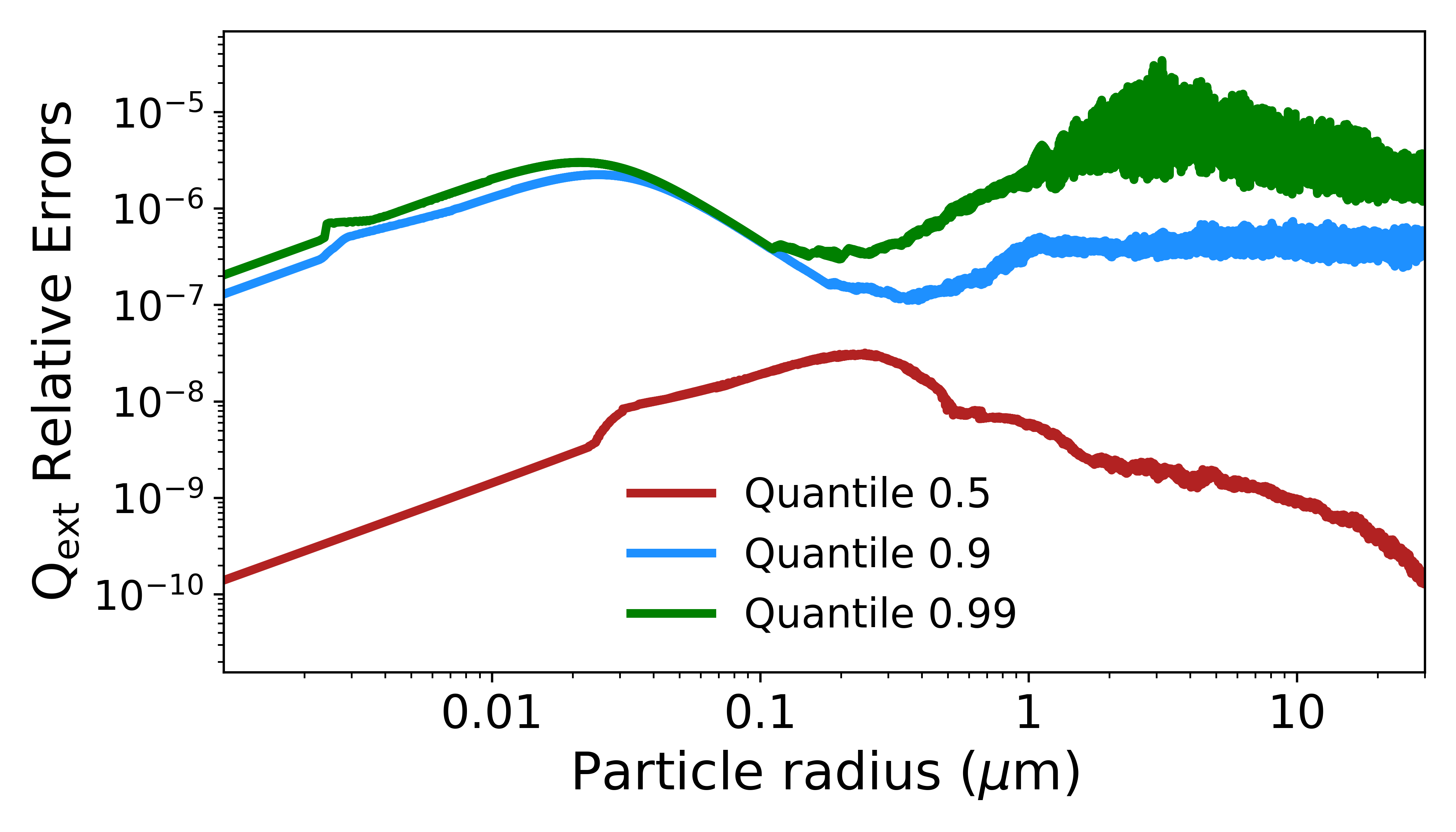}
    \caption{$\rm{Q_{ext}}$ relative error between \textsc{PyMieScatt} and linear interpolation for each radius interval in the Titan tholin grid. For each radius, the $\rm{Q_{ext}}$ is computed for 517 wavelengths from $0.3$ to $50$ microns. The $0.5$, $0.9$ and $0.99$ quantile shown in red, blue and green respectively are computed at each radius from the relative error versus wavelength array. }
    \label{FigQextErr}
\end{figure}

\begin{figure*}[h!]
    \centering
    \includegraphics[width=\textwidth]{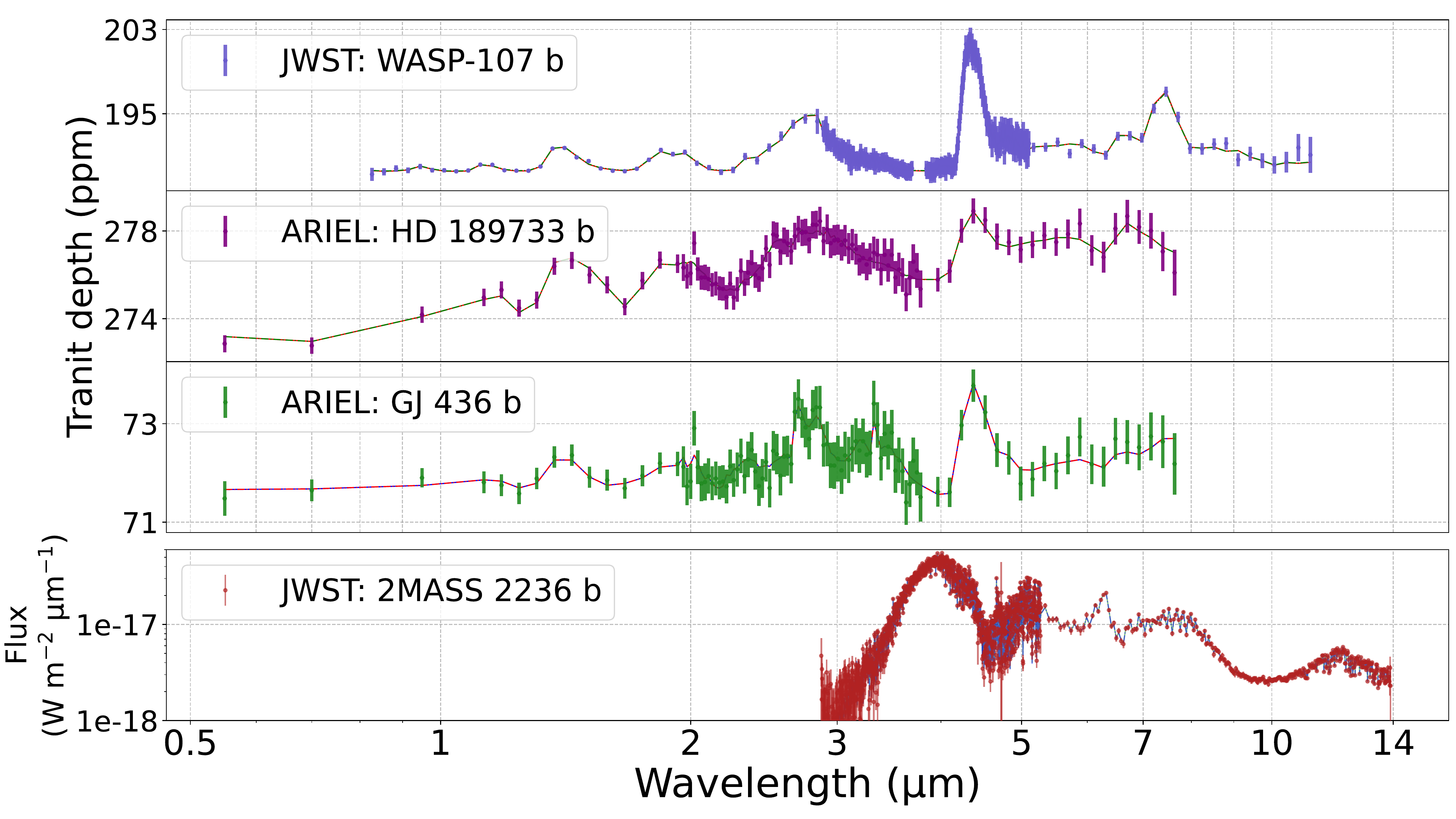}
    \caption{Synthetic spectra created for the self-retrievals, inspired by WASP\,107\,b, HD\,189733\,b, GJ\,436\,b and 2MASS\,2236\,b. To generate the errorbars for WASP\,107\,b we use the \textsc{PandExo} framework \citep{batalha_PandExoCommunityToolTransiting_2017}. The errorbars for 2MASS\,2236\,b come from real JWST observations (PID:1188). For ARIEL we make use of ARIELRAD to simulate tier 3 noise, stacking 2 transits for HD\,189733\,b and 10 for GJ\,436\,b. In each panel, the colored lines represent the best fit models using \textsc{TauREx-PyMieScatt} and our grid based methods.}
    \label{fig:spectra}
\end{figure*}

We start by computing the largest grid step $\delta a$ that provides an accuracy lower than $\epsilon$ for the maximum ($a_{max} = 30\,\mu$m) and minimum ($a_{min} = 1$ nm) radii, for example in the case of the tholins, $\delta a_{min} = 0.05$ nm and $\delta a_{max} = 1$ mm. One should remind that for each particle radius, the $\rm{Q_{ext}}$ is computed over a wavelength range with $\lambda\in[0.3, 50]\,\mathrm{\mu m}$. Thus, for at a given radius, the accuracy is in fact a distribution of values: one for each wavelength. We impose that, for each $a$, the $90$ quantile of the accuracy should be inferior to $\epsilon$. Using these grids steps we compute the linear coefficient $m$ and $b$  so that the equation $\delta a = m \cdot a + b$ verifies $\delta a_{min} = 0.05$ nm and $\delta a_{max} = 1$mm for $a \in [0.001, 30]\,\mu$m. Using this formula for $\delta a$ we compute the radius grid $\mathcal{E}_a^0$. Then, for each radius $a$ in $\mathcal{E}_a^0$ we compute the accuracy. If the accuracy is superior to $\epsilon$ for a radius $a_{fail}$ we repeat the above procedure between $a_{min}$ and $a_{fail}$. We thus obtain $\{m_1,b_1\}$ for $a\in[a_{min}, a_{fail}]$ and $\{m_2,b_2\}$ for $a\in[a_{fail}, a_{max}]$ and can compute the radius grid $\mathcal{E}_a^1$. We repeat the process until the accuracy is reached for the full radius range (between one and four times for the species in Table \ref{tab:placeholder}).

Using the method described above, we built a grid of $\rm{Q_{ext}}$ for each species. We consider the species shown in Table \ref{tab:placeholder}. As an example, for Titan's tholins, a grid of $58\,000$ radii points is necessary to cover $1$ nm to $30\,\mu$m, and requires 240\,Mb of memory space. In Fig. \ref{FigQextErr} we show the error introduced by our linearisation of the Mie theory in the case of the Titan tholins. Our grid strategy ensures that each radius should have more than $90\%$ of the $\rm{Q_{ext}}$ relative error below the accuracy limit of $10^{-4}$ (see Fig. \ref{FigQextErr}), which should be {\it indistinguishable} in current JWST and Ariel retrievals but provides significant speed-ups.

\subsection{Scattering coefficient and asymmetry parameter}\label{sec:Qscat_g}

\noindent Although we primarily focus on $\rm{Q_{ext}}$, Mie theory is also used to compute the scattering efficiency $\rm{Q_{scat}}$ and asymmetry parameter $\rm{g}$.
While $\mathrm{Q_{scat}}$ and $\mathrm{g}$ are not currently included in free atmospheric retrieval frameworks, they are commonly required by self-consistent models. Consequently, eliminating on-the-fly Mie calculations in the general case requires pre-computing all three quantities. As $\rm{Q_{scat}}$ and $\rm{g}$ are also computed with the Mie coefficients $a_n$ and $b_n$, \textsc{PyMieScatt} computes the three coefficients all at once:

\begin{equation}
    \mathrm{Q}_{\text{scat}}(a, \lambda) = \frac{2}{x^2} \sum_{n=1}^{n_{max}} (2n + 1)  \left( | a_n(m(\lambda), x)|^2 + |b_n(m(\lambda), x)|^2 \right)
\label{eq:2}
\end{equation}
and 

\begin{equation}
    \rm{g}(a) = \frac{1}{2} \int_{0}^{\pi} p(a, \alpha)\rm{cos}(\alpha)\rm{sin}(\alpha)\rm{d}\alpha
\label{eq:3}
\end{equation}

where $\alpha$ is the scattering angle and $p(a,\alpha)$ the scattering phase function. Please refer to \cite{kitzmann_optical_2018} for the detailed expression of  $p(a,\alpha)$ and \cite{bohren_absorption_1983} for the full derivation. Thus, for each of our $\rm{Q_{ext}}$ grids we have also computed one grid for $\rm{Q_{scat}}$ and another for $\rm{g}$. These grids utilize the same radius grids as for  $\rm{Q_{ext}}$, so we remind users that they might not be as optimal (i.e., we did not impose a tolerance criteria for $\rm{Q_{scat}}$ and  $\rm{g}$). Nonetheless, performing the same error analysis as for $\rm{Q_{ext}}$, we find that the accuracy is similar, with our grid most of the time satisfying our tolerance criteria of $\epsilon = 10^{-4}$. For $\rm{Mg_2SiO_4}\, amorp\,sol-gel$, $\rm{MgSiO_{3}\, amorph\,glass}$, $\rm{MgSiO_{3}\,amorph\,sol-gel}$ and the $\rm{Titan\,tholins}$ the relative error on $\rm{Q_{scat}}$ were superior to $\epsilon$ for the very small particle radii i.e., $a < 10$ nm. To correct this, we decreased the radius grid step in this region for those four species. Adding only $\approx 500$ more radius steps, the accuracy of the four $\rm{Q_{scat}}$ grids listed above increased passed our threshold of  $\epsilon = 10^{-4}$. The  $\rm{Q_{scat}}$ and $\rm{g}$ grids are also available on the Zenodo\footnote{\url{https://doi.org/10.5281/zenodo.17456673}} and share the same memory size as their  $\rm{Q_{ext}}$ counterparts (see Table \ref{tab:placeholder}). However, since \textsc{TauREx} does not use $\rm{Q_{scat}}$ and $\rm{g}$, we do not test these grids further in the self-retrieval verification process described below.\\

\subsection{Synthetic spectra}

Our methodology ensures relative errors on $\rm{Q_{ext}}$ below $0.1\%$ for all the species. However, radiative transfer is a highly nonlinear process with respect to opacities (they impact is via the Beer Lambert exponential law), even negligible errors in the $\rm{Q_{ext}}$ and linearizations can introduce significant differences in the final spectrum (and retrieved parameters since the sampled space is also non-linear). To assess the difference of the \textsc{TauREx-PCQ} approach for aerosol versus full Mie (i.e., \textsc{PyMieScatt}) retrievals, we conduct four different tests (see Table  \ref{tab:cases}). We consider standard scenarios---in transit and emission spectroscopy by direct imaging---with JWST and with the upcoming Ariel Telescope \citep{tinetti_ariel_2021}. Each test case we create is strongly inspired by a real exoplanet, but we do not claim here that the simulations presented here are actually relevant for these particular exoplanets. We study atmospheres similar to WASP-107 b, GJ 436 b, HD 189733\,b and 2MASS J2236+4751\, b (hereafter 2M2236\,b) \citep{dyrek_SO2SilicateCloudsNo_2024, changeat_CloudHazeParameterizationAtmospheric_2025,mukherjee_JWSTPanchromaticThermalEmission_2025,moses_DISEQUILIBRIUMCARBONOXYGENNITROGEN_2011,bowler_PlanetsLowmassStarsPALMS_2017}. To simulate these cases, we use the code \textsc{TauREx}. The parameters used for the simulations are summarized in Table \ref{tab:planet_gas_params} $\&$ Table \ref{tab:cloud_params}. For the JWST WASP\,107\,b synthetic spectra, we utilize \textsc{PandExo} to simulate the errorbars \citep{batalha_PandExoCommunityToolTransiting_2017}. We use the uncertainties from the real observations for 2M2236\,b since this information is readily available (PID:1188). For Ariel, we employ the radiometric model ArielRad\footnote{For reference, Ariel simulations employ: ArielRad v2.4.26, ExoRad v2.1.111, Payload v0.0.17.} to simulate realistic Tier 3 noise in the case of GJ 436\,b and HD 189733\,b \citep{mugnai_ArielRadArielRadiometricModel_2020}. The four simulated observations, obtained after convolving the atmospheric model with the instrument noise profiles, are shown in Fig \ref{fig:spectra}. The retrievals with the best fit model for the on-the-fly Mie calculation versus our \textsc{TauREx-PCQ} approach are also shown and discussed in the next section. In every case, the input simulation is scattered using the standard deviation from the noise profiles (i.e., we assume Gaussian noise and utilize a single instance).

\section{Results}\label{sec:3}
\subsection{JWST transmission spectroscopy WASP 107\,b}

WASP\,$107$\,b is a warm ($\rm{T_{eff}} = 740$ K) Neptune with an inflated radius of $\approx 0.84\,\rm{R_J}$ which shows signs of photochemistry, significant mixing, dis-equilibrium chemistry and a $10\,\mu$m silicate absorption feature. This absorption signature at $10\,\mu$m is created by the stretching of Si--O atomic bonds in aerosols. The family of molecules that contributes to this feature---namely the silicates---regroups among other SiO, SiO$_2$, MgSiO$_3$, Mg$_2$SiO$_4$ or olivine (MgFeSiO$_4$). These particles condense with different crystalline structures, for instance "amorphous" or "alpha" for SiO$_2$, depending on the formation conditions. Different structures of the particles will yield different optical indexes and change the shape of the resonance bands. For WASP\,$107$\,b we chose to simulate the atmosphere using MgSiO$_3$ in its "amorphous glass" form. The atmospheric parameters used to create the synthetic spectra, summarized in Table \ref{tab:planet_gas_params} $\&$ \ref{tab:cloud_params}, are inspired from \cite{changeat_CloudHazeParameterizationAtmospheric_2025}. We build our synthetic spectrum with \textsc{TauREx} including errorbars from JWST NIRISS, NIRSpec $\mathrm{G395H}$ and MIRI LRS observation and scattering the data points. The $\rm{MgSiO_3\, amorph\,glass}$ cloud layer has particles with a $0.5\rm{\mu m}$ radius. In Fig. \ref{Fig:wasp} we show the posterior distribution of the same retrieval performed with on-the-fly Mie computation and \textsc{TauREx-PCQ}. The retrieved parameters and posterior distributions between the two methods are the same, which validates our approach for JWST transits of this quality. We record computation time, and the \textsc{TauREx-PCQ} retrieval was $1.4$ times faster than the retrieval using \textsc{PyMieScatt}. \\
However, when including more species, the two methods scale differently. This is because of the $a$ dependence of compute times with \textsc{PyMieScatt}, which increase with larger $a$ as described previously, and significantly penalize retrievals with a larger number of species (i.e., sampled $a$) and live points. With pre-computed $\rm{Q_{ext}}$ grids, having several cloud species does not significantly change the computational time. When including 4 condensed species (i.e., SiO, SiO$_2$, MgSiO$_3$, Mg$_2$SiO$_4$) in a similar WASP-107\,b scenario, the \textsc{TauREx-PCQ} retrieval is $17$ times faster.

\subsection{Ariel transmission spectroscopy HD 189733\,b and GJ\,436\,b}

By observing $\sim 1000$ exoplanet atmosphere, the Ariel mission will be able to study clouds and aerosols of exoplanets on a population scale. As a complementary example to the JWST simulations, and to highlight the relevance of aerosol modeling with Ariel, we also explore simulations with this future observatory. We chose to simulate two synthetic {bright} exoplanets inspired by the hot Jupiter HD\,189733\,b for which $\rm{SiO_2}$\footnote{we here include the amorphous glass form of  $\rm{MgSiO_3}$} could be present, and the sub-Neptune GJ\,436\,b for which aerosols similar to the tholins on Titan could explain the blueward Mie Scattering slope. We use atmospheric parameters from \cite{zhang_RetrievalsNIRCamTransmissionEmission_2024} and \cite{inglis_QuartzCloudsDaysideAtmosphere_2024} for HD\,189733\,b and \cite{mukherjee_JWSTPanchromaticThermalEmission_2025} for GJ\,436\,b to build our synthetic spectrum with \textsc{TauREx} (see Table \ref{tab:planet_gas_params} $\&$ \ref{tab:cloud_params}). Then, we utilize \textsc{ArielRad} to simulate the noise for a transit spectrum, assuming that the target is observed in Tier 3 resolution (i,e., native Ariel resolution), stacking 2 and $10$ transits for HD\,189733\,b and  GJ\,436\,b respectively  \citep{mugnai_ArielRadArielRadiometricModel_2020}. The posterior distributions for both planets are shown in Fig \ref{Fig:HD} and \ref{Fig:GJ}. As for the previous WASP\,$107$\,b JWST case, each figure shows the on-the-fly PyMieScatt retrieval versus the \textsc{TauREx-PCQ} retrieval. As in the JWST case, the retrieved parameters for HD\,189733\,b and GJ\,436\,b with Ariel are independent from the aerosol model we use, validating that our $\rm{Q_{ext}}$ grid approach is numerically converged and equivalent for Ariel data. Similar speed improvements are observed: the \textsc{TauREx-PCQ} retrieval is $2.3$ times faster for  HD\,189733\,b and $2.75$ faster for GJ\,436\,b. For the Ariel survey concept, speed-ups of this order are crucial to ensure analysis of hundreds of atmospheric spectra. In addition, even though the silicates main absorption feature (at $\lambda \sim 9 - 10\,\mu$m) is outside the wavelength range of Ariel ($0.6 - 7.9\,\mu$m), if priors on the type of particles can be placed---for instance via synergy with JWST-MIRI \citep{changeat_synergetic_2025} or using more informative aerosol formation models \citep{ma_YunMaEnablingSpectralRetrievals_2023}---Ariel could able to inform us on the particle properties (such as radius, altitude, and cloud density, see Fig. \ref{Fig:HD}).

\subsection{JWST emission spectroscopy 2MASS J2236+4751\,b }

While the field of transiting exoplanets is delivering 10\,s of hight quality atmospheric spectra, observations of brown dwarf, planetary mass companion and exoplanet in direct imaging with JWST have yielded much higher resolution ($R > 2700$) and signal-to-noise ratio ($SNR > 100$) spectra \citep{miles_JWSTEarlyreleaseScienceProgram_2023,barrado_15NH3AtmosphereCoolBrown_2023}. Currently, these datasets are numerically the most challenging for atmospheric retrievals. Interpreting datasets with such information content challenges our analysis framework in terms of complexity, model accuracy and convergence times. In particular, a number of these spectra exhibits clear absorption features from condensate species: for instance, L dwarfs have strong silicate absorption features \citep{miles_JWSTEarlyreleaseScienceProgram_2023, molliere_EvidenceSiOCloudNucleation_2025}. For such objects, very few Bayesian atmospheric retrievals have been attempted as robust convergence can only be achieved over several months of computations in HPC facilities. Yet, characterizing the clouds of such objects---particularly in the L/T transition---is essential to understand the mechanisms that shape the atmosphere of L and T dwarfs. L type objects are notorious for their silicates absorption features, but determining the specific particles creating this feature is a tedious and time consuming process \citep{molliere_EvidenceSiOCloudNucleation_2025}. Significantly decreasing the computational cost of the clouds model is required to enable more widespread retrievals analysis and to test several aerosol hypothesis in reasonable time frames. We validate the benefits of our \textsc{PCQ} approach on a simulated high information content spectrum in direct imaging. We create a synthetic spectrum inspired from the late L planetary mass companion 2MASS J2236+4751\,b with an added $\rm{MgSiO_3}$ cloud layer \citep{bowler_PlanetsLowmassStarsPALMS_2017} (see specific input parameters in Table \ref{tab:planet_gas_params} $\&$ \ref{tab:cloud_params}). We use errorbars from the real planet's observation with NIRSpec IFU and MIRI LRS (PID:1188). The posterior distribution of the retrievals using \textsc{TauREx-PyMieScatt} and  \textsc{TauREx-PCQ} are shown in Fig. \ref{Fig:2M}, demonstrating that both methods accurately retrieve the appropriate solution. Hence, using linearized $\rm{Q_{ext}}$ with $\epsilon < 10^{-4}$ is appropriate for even the highest information content spectra available today. We note an improvement of $2.1$ in the convergence speed for this case.

\section{Summary and conclusions}

In this work, we present a new approach for incorporating aerosols into atmospheric retrievals that substantially reduces their computational cost. Aerosols are typically modeled by computing extinction, scattering coefficients and asymmetry parameter from the complex refractive indices of condensate species by solving the Lorenz–Mie equations for a range of particle radii and wavelengths. However, the repeated evaluation of Mie theory during a free retrieval or self-consistent model is computationally expensive and often dominates run time. To alleviate this bottleneck, we constructed pre-computed extinction, scattering and asymmetry grids for seven aerosol species (silicates and the Titan tholins) covering a wide range of particle radii and wavelengths. This strategy is effectively equivalent to linearising Mie theory with respect to particle size. We quantified the interpolation errors introduced by this approximation and evaluated their impact on retrieval performance. We also focus on the optical constants which are required for Mie theory computations. Indeed, the wavelength resolution used for these constants can impact the retrieval outcomes for very high information datasets such as direct imaging. We show that using an adaptive wavelength grid for species with sharp features is required to avoid systematic errors. For the $\mathrm{Q_{ext}}$ grids we employ self retrievals on synthetic spectra based on four representative atmospheric scenarios, each scattered with real or realistic errorbars, we find that the interpolation errors are negligible and do not affect the retrieved atmospheric parameters. By removing on-the-fly Mie computations, retrievals converge significantly faster. The speed-up depends on the true particle size and on the number of aerosol species included, with gains of 1.4–2.3 for single-species cases. A key advantage of our method is its scaling with the number of clouds: while the Mie theory approach slows down exponentially as more condensates are added, the \textsc{TauREx-PCQ} method runtime is essentially independent of the number of clouds. For instance, although a single-cloud retrieval using $\mathrm{Q_{ext}}$ grids achieved a speed-up of 1.4, the same retrieval with four clouds became 17 times faster than the corresponding retrieval using direct Mie calculations. Most importantly, these significant speed-ups are achieved without any loss in accuracy. We make these $\mathrm{Q_{ext}}$, $\mathrm{Q_{scat}}$ and $\mathrm{g}$ grids freely available for the community at \url{https://doi.org/10.5281/zenodo.17456673}. We also publish  \textsc{TauREx-PCQ}\footnote{The code and documentation can be found at : \url{https://github.com/groningen-exoatmospheres/taurex-PCQ}} a plugin that integrates the $\rm{Q_{ext}}$ grids in the  \textsc{TauREx} retrieval framework. This new method for clouds in atmospheric retrieval is essential to enable the retrieval analysis of high information content datasets within reasonable timescales. Furthermore, it is also a key development to enable the uniform analysis of large exoplanet populations.

\begin{acknowledgements}
M. Voyer acknowledges funding support by CNES. The authors thank P-O. Lagage for his precious inputs and insightful discussions. This project was provided with computing HPC and storage resources by GENCI at TGCC thanks to the grant 2024-15722 and 2025-15722 on the supercomputer Joliot Curie’s SKL and ROME partition. This publication is part of the project ``Interpreting exoplanet atmospheres with JWST'' with file number 2024.034 (PI: Changeat) of the research programme ``Rekentijd nationale computersystemen'' that is (partly) funded by the Netherlands Organisation for Scientific Research (NWO) under grant \url{https://doi.org/10.61686/QXVQT85756}. This work used the Dutch national e-infrastructure with the support of the SURF Cooperative using grant no. 2024.034. 
\end{acknowledgements}

%
%

\bibliographystyle{aa} 
\bibliography{refs} 

\vfill
\clearpage

\appendix
\onecolumn
\section{Appendix: additional figures}\label{Appendix:obs_params}

\begin{figure}[ht!]
   \centering
    \includegraphics[width=0.99\textwidth]{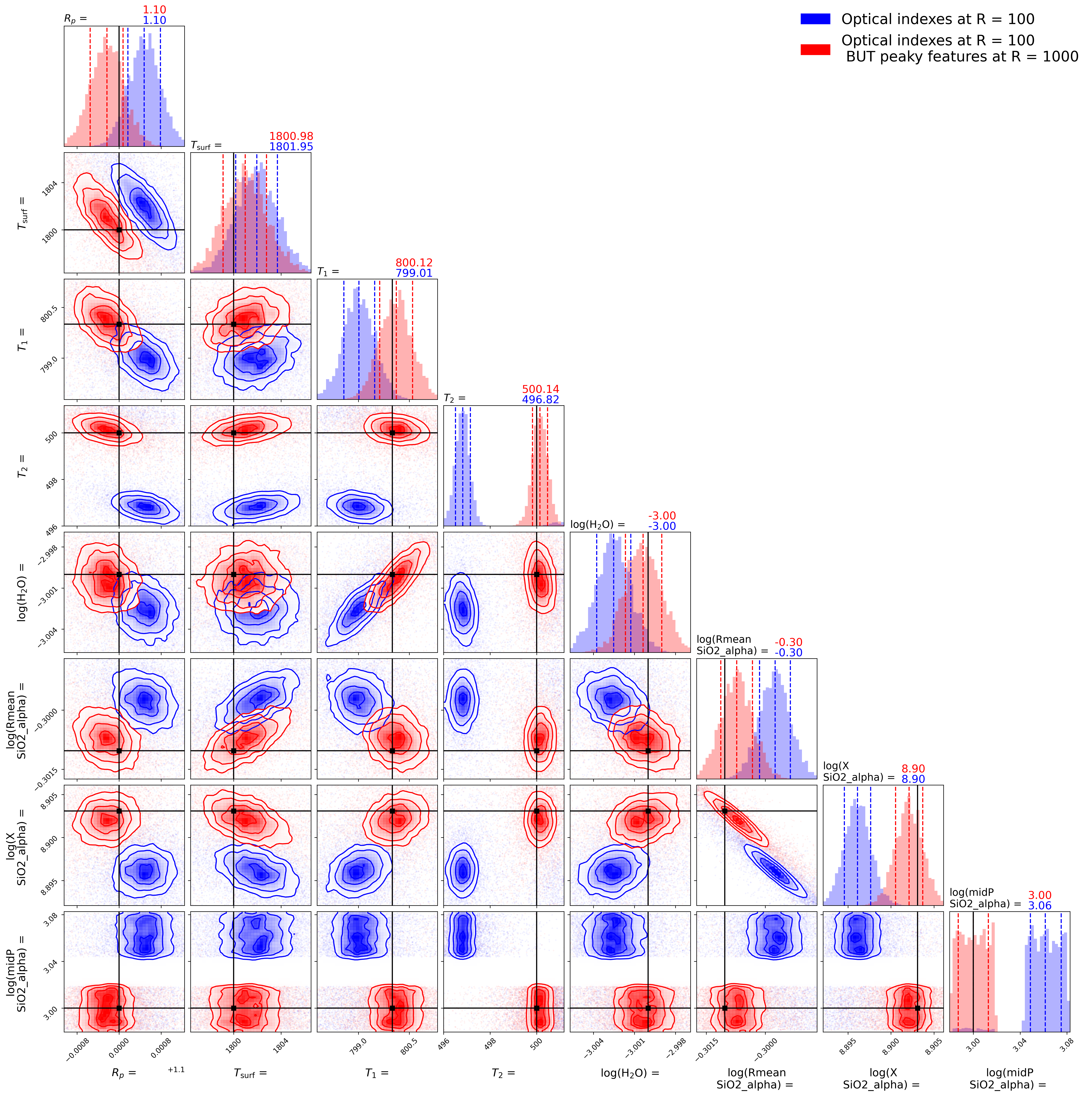}
   \caption{Posterior distributions for two retrievals on a 2M $2236$\,b inspired planet with a SiO$_2\,\alpha$ absorption feature. Truths are shown in black. The blue retrieval uses optical constants at resolution 100 whereas the red retrieval utilize a mix $100$ and $1000$ resolution for the optical constants.}
              \label{Fig:cornerResolution}
\end{figure}

\begin{figure*}[ht!]
   \centering
    \includegraphics[width=0.99\textwidth]{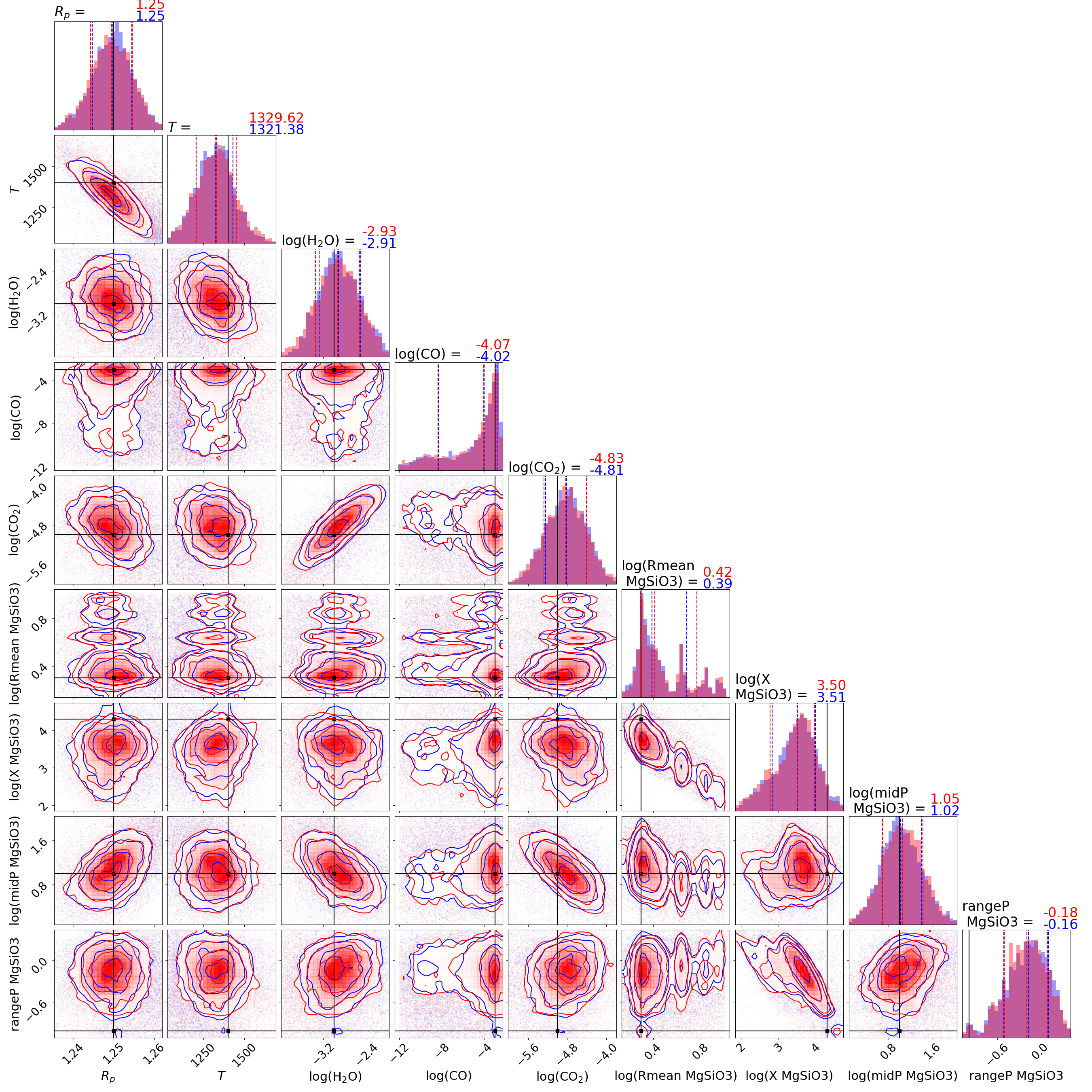}
   \caption{Same as Fig. \ref{Fig:wasp}, but for HD\,189733\,b.}
              \label{Fig:HD}%
\end{figure*}

\begin{figure*}[ht!]
   \centering
    \includegraphics[width=0.99\textwidth]{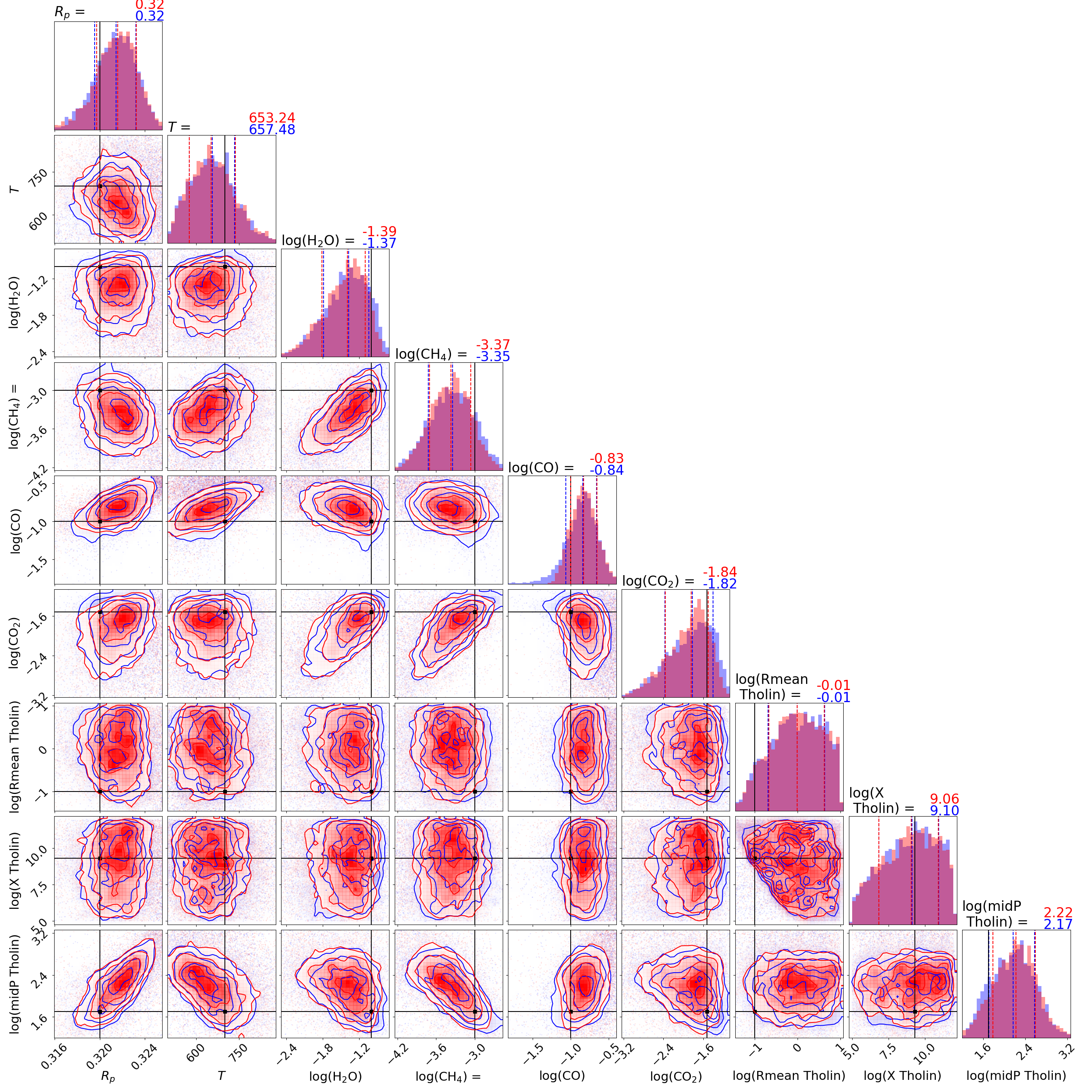}
   \caption{Same as Fig. \ref{Fig:wasp}, but for GJ\,436\,b.}
              \label{Fig:GJ}%
\end{figure*}

\begin{figure*}[ht!]
   \centering
    \includegraphics[width=0.99\textwidth]{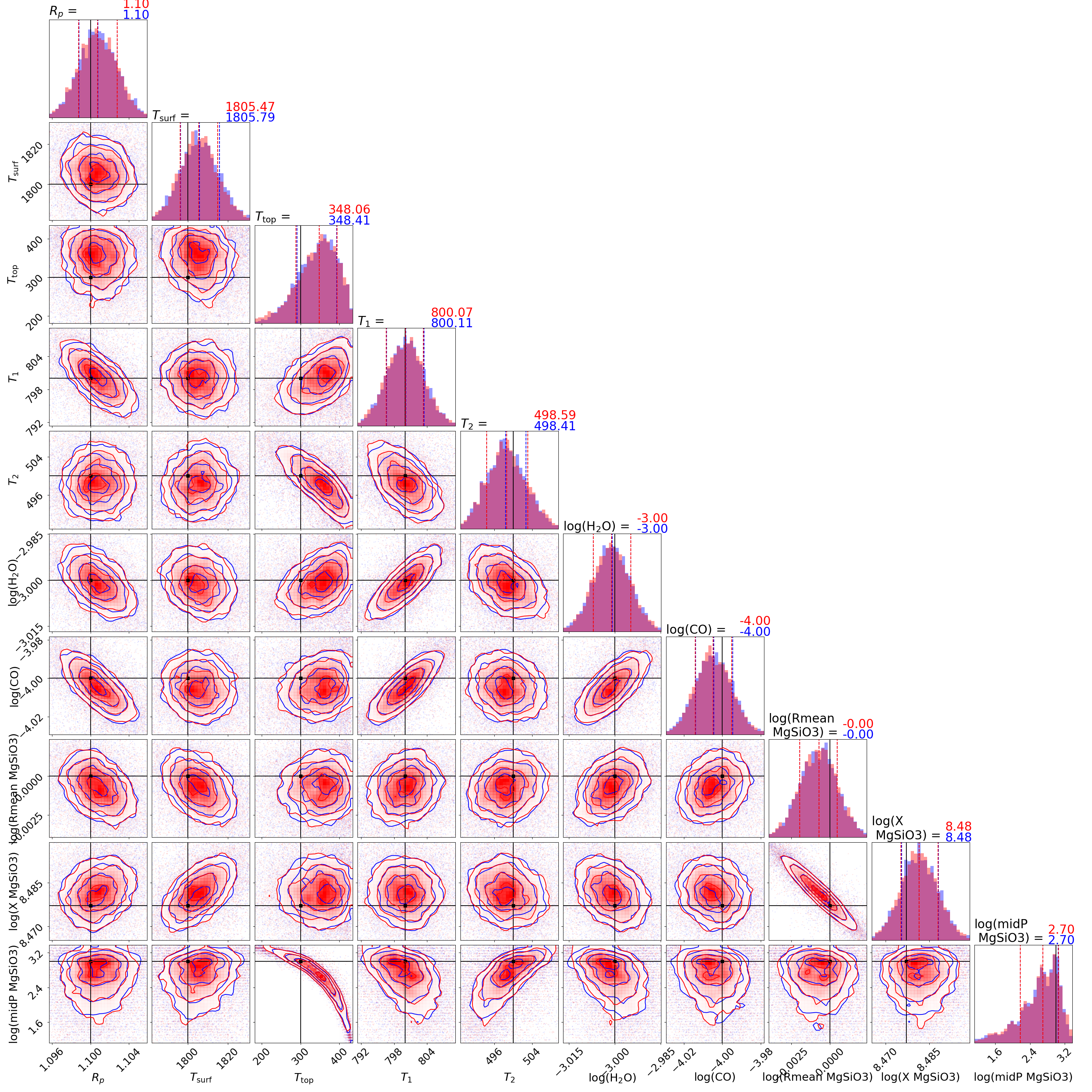}
   \caption{Same as Fig. \ref{Fig:wasp}, but for 2M\,2236\,b.}
              \label{Fig:2M}%
\end{figure*}

We present Fig. \ref{Fig:cornerResolution} which shows the impact of the resolution of the optical indexes on the retrieved parameters for a 2M $2236$\,b inspired planet with an added SiO$_2\,\alpha$ absorption feature. Then, we show the posterior distributions for the HD 189733 b, GJ 436 b and 2M 2236 b inspired self-retrieval respectively in Fig. \ref{Fig:HD},  Fig.\ref{Fig:GJ} and Fig. \ref{Fig:2M}.

\newpage
\,
\newpage
\,
\newpage

\section{Appendix: additional tables}\label{Appendix:obs_params}

The following Tables \ref{tab:planet_gas_params} \& \ref{tab:cloud_params} summarize the atmospheric parameters used to create the synthetic spectra shown in Fig. \ref{fig:spectra}.

\begin{table}[ht]
    \centering
    \caption{Parameter used for generate the synthetic spectra $1/2$.}
    \begin{tabular}{lccccccccl}
        Planet & $\mathrm{R_P}$ [R$_{\mathrm{J}}$]& $\mathrm{M_P}$ [M$_{\mathrm{J}}$] & SMA [AU]& $\mathrm{T}$ [K]& H$_2$O & CO$_{2}$ & CO & CH$_{4}$  &SO$_2$\\ \hline\hline
        WASP-107\,b & 0.85& 0.12 & 0.055 & 750 & $1\times10^{-3}$& $1\times10^{-4}$& $5\times10^{-2}$& n/a  &$2\times10^{-6}$\\
        HD\,189733\,b & 1.25 & 1.13 & 0.031 & 1400 & $1\times10^{-3}$ & $1\times10^{-5}$ & $5\times10^{-3}$& n/a  &n/a\\
        2M\,2236\,b& 1.10 & 12 & n/a & $1800, 900, 500, 300$ & $1\times10^{-3}$ & n/a & $1\times10^{-4}$ & n/a  &n/a\\
        GJ\,436\,b & 0.32 & 0.05 & 0.020 & 700 & $1\times10^{-1}$ & $3\times10^{-2}$& $1\times10^{-1}$& $1\times10^{-3}$  &n/a\\
        \hline
    \end{tabular}
    \label{tab:planet_gas_params}
\end{table}

\begin{table}[ht]
    \centering
    \caption{Parameter used for generate the synthetic spectra $2/2$.}
    \begin{tabular}{lcccccc}
        Planet & Cloud species & Particle radius [$\mu$m] & $\mathrm{P_{mid}}$ [Pa] & $\Delta P$ [Pa] & $X_{\mathrm{cloud}}$ \\ \hline\hline
        WASP-107\,b & MgSiO$_3$ (am.) & 0.5 & $2\cdot10^{2}$ & 1.0 & $5\times10^{6}$ \\
        HD\,189733\,b & MgSiO$_3$ (am.) & 2.0 & $10^{1}$ & 0.1 & $2\times10^{4}$ \\
        2M\,2236\,b & MgSiO$_3$ (am.) & 1.0 & $10^{3}$ & 0.1 & $3\times10^{8}$ \\
        GJ\,436\,b & Titan tholin & 0.1 & $10^{2}$ & 0.1 & $1\times10^{9}$ \\
        \hline
    \end{tabular}
    \tablefoot{
        $\bar{r}_{\mathrm{cloud}}$ is the mean particle radius of the condensate, 
        $P_{\mathrm{mid}}$ is the pressure at which the cloud optical depth peaks, 
        $\Delta P$ is the pressure range (in dex) of the cloud layer, 
         and $X_{\mathrm{cloud}}$ is the assumed particle mixing ratio.
    }
    \label{tab:cloud_params}
\end{table}

\end{document}